\documentclass{netsci-project}

\usepackage[
    backend=biber,
    style=authoryear,
    natbib=false,
    maxcitenames=2,
    minbibnames=1, maxbibnames=99, 
    url=false, 
    isbn=false,
    eprint=false,
   doi=true,
    ]{biblatex}

\usepackage{xcolor}
\usepackage{todonotes}
\usepackage{comment}
\usepackage{lipsum}
\usepackage[english]{babel}
\usepackage[letterpaper,top=2cm,bottom=2cm,left=3cm,right=3cm,marginparwidth=1.75cm]{geometry}
\usepackage{amsmath}
\usepackage{csquotes}
\usepackage{graphicx}
\usepackage{multicol}
\usepackage[hypcap=false]{caption}
\usepackage{import}
\usepackage{float}
\usepackage[colorlinks=true, allcolors=blue]{hyperref}
\usepackage[ruled,vlined]{algorithm2e} 

\subjectarea{Blockchain \& Distributed Ledger Technologies}
\newenvironment{Figure}
  {\par\smallskip\noindent\minipage{\linewidth}}
  {\endminipage\par\smallskip}

\begin{document}
\firstpage{1}

\title{Epidemics on Networks}
\author{Jan Kreischer, Adrian Iten, Astrid Jehoul}
\course{Network Science}
\school{Faculty of Business, Economics and Informatics}
\date{17.12.2021}

\maketitle

\begin{multicols}{2}[]
\todo{Finish abstract}
\begin{abstract}
\textit{Despite centuries of work on containment and mitigation strategies, infectious diseases are still a major problem facing humanity. This work is concerned with simulating heterogeneous contact structures and understanding how the structure of the underlying network affects the spread of the disease. For example, it has been empirically demonstrated and validated that scale free networks do not have an epidemic threshold. Understanding the relationship between network structure and disease dynamics can help to develop better mitigation strategies and more effective interventions.}
\end{abstract}

\section{Introduction}

The vulnerability of networks to the spread of viruses has been painfully illustrated and manifested several times in recent years. Computer viruses and ransomware, such as WannaCry and NotPetya, autonomously spreading across the internet, encrypting and halting the IT-systems of entire companies, became major problems in 2017. However, the problem of virus diffusion is not limited to computer networks. At the end of 2019, the world was hit by a different type of virus - SARS-CoV-2. The virus can be transmitted from person-to-person, potentially causing a severe respiratory disease (COVID-19), and has been able to spread rapidly across all continents, bringing entire countries to a standstill.\\

In our final project we want to assess how viruses like SARS-CoV-2 spread on social networks. Hereby, nodes represent agents and the edges represent potentially infectious interactions between agents. At any moment in time, the disease might be transmitted between individuals who are connected via an edge. For modelling an epidemic on networks, many different types of models exist like compartmental models, agent-based models amongst other. However, most models are based on the assumption that a contact between two individuals happens with equal probability, the so called homogeneous mixing assumption. This assumption simplifies the calculation, but often does not hold in reality (\cite{AlexanderKaraivanov2020}).\\

Agent-based models have greater flexibility and allow the simulation of heterogeneous contact structures. The effects that the structure of the underlying network might have on the spread of the disease were investigated. For this purpose three, different synthetic network models and three 
three real social networks will be used. Wherever possible, results are compared with real-world observations. By better understanding the spread of diseases in networks, more effective intervention strategies (like social policies or targeted lockdowns) can be developed. 

\section{Theory}

A short overview of the theoretical concepts is given in this section. This project focuses on simple contagions, where an individual can get infected after a single exposure to the virus. This differs from complex contagions, where an agent has to be exposed multiple times before an infection can occur.\\

Most models for epidemiology incorporate the homogeneous mixing assumption, meaning all, potentially infectious, encounters between agents are equally likely to happen. Accordingly, every agent gets in contact with any other agent with an equal probability. However, this simplifying assumption is often not realistic. Therefore, agent based simulations might be more suitable, since they are able to deal with contact heterogeneity (\cite{ShwetaBansal2007}).\\

A key parameter in epidemiology is the basic reproduction number $R_0$. This is the expected number of secondary cases, caused by a single infected individual that is placed into a fully susceptible population. It can be estimated at the start of an epidemic, when the whole population is susceptible, before widespread immunity or vaccination. The number reveals a lot about the expected disease dynamics. If $R_0$ is greater than 1, each existing infection causes more than one new infection, potentially leading to an exponential spread of the disease. If not contained, this might lead to an outbreak or epidemic. According to what was discussed in class, COVID-19 is estimated to have a $R_0$ between 2 and 4. In contrast, if $R_0$ is below 1, each existing infection will cause less than one new infection. In this case, the disease spread will decline and eventually die out. In reality, a population will rarely be totally susceptible to a virus, since prior infection or vaccination has created a certain degree of immunity. Therefore, the average number of secondary cases caused by a single infection will be lower. The effective reproductive number ($R_e$) denotes the average number of secondary cases generated in a population made up of both susceptible and non-susceptible agents.\\

\subsection{Compartmental Models}
Compartmental models (CMs) are the most common and general technique for modelling infectious diseases. CMs describe the transition of agents between discrete compartments over time, depending on their infection status (\cite[p.3]{gallagher2017comparing}). Internal details describing the complete state of an individual agent are unimportant, as we've seen during the course. Furthermore, simple assumptions about movements between these compartments are used e.g. a closed population is assumed, meaning no immigration or emigration happens. Over time, these models have grown more sophisticated, incorporating the contact structure of the population more precisely and taking advantage of increased computational resources. Nevertheless, these models are still limited regarding the complexity they can represent.\\

\textbf{Control Parameters} are given as an input to the simulation (e.g network type, infection rate) and specify the rules of the mechanism. They are adjusted in order to understand their effect on the emergent system properties.
\begin{itemize}
    \item \textit{Infection rate} $\beta$: The risk/probability of an infection happening during an interaction between a susceptible and infected agent. It can be changed through interventions like social distancing or wearing masks.
    \item \textit{Recovery rate} $\gamma$: The inverse of the period of infection. Each infected person runs through the course of his sickness, and finally is removed from the number of those who are sick, by recovery or by death (\cite[p.701]{kermack1927contribution}). It is difficult to change the recovery rate if no medication/vaccination is possible.
    \item \textit{Waning immunity rate} $\alpha$: The rate at which, agents become susceptible to reinfection, after recovery or vaccination, due to a progressive decline of protective antibodies. 
\end{itemize}

\begin{Figure}
\centering
\includegraphics[width=\linewidth]{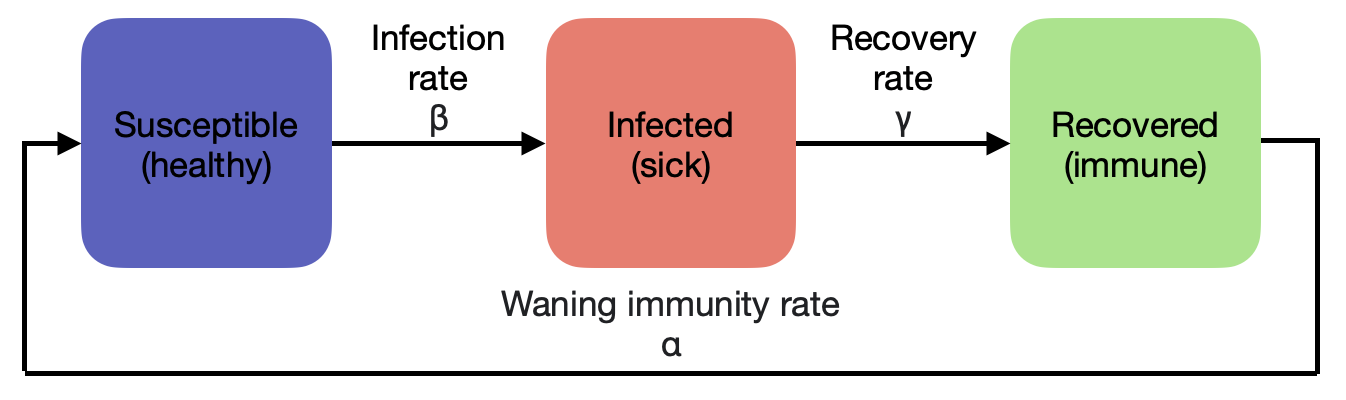}
\captionof{figure}{Simplified SIRS model visualization}
\end{Figure}
\textbf{Order parameters}, are measured throughout the simulation and quantify the emergent state of the system.

\begin{itemize}
    \item \textit{Total population size} $N$: Here assumed to be constant, implying a mortality rate $\mu$ of 0.
    \item \textit{Number of susceptible individuals} $N_S$ 
    \item \textit{Number of infected individuals} $N_I$
    \item \textit{Number of recovered individuals} $N_R$
\end{itemize}

\subsubsection{Susceptible Infected (SI) Model} \hfill \break
The SI model is the simplest compartmental models (CMs). In this case, the population is divided into only two compartments: an agent is either Susceptible (healthy) or Infected (sick). The probability that a susceptible agent becomes infected is given by the infection rate per contact $\beta$ multiplied by the number of social contacts per time unit. All contacts have the same probability of infecting an agent and all contacts can be considered to be independent processes. What happens in one time interval is independent of what happens in the next time interval. Under this assumption, the whole population will eventually become infected. 

\subsubsection{SIR Model} \hfill \break
The SIR model is an extension of the SI model. The population is divided into 3 compartments namely: susceptible (S), infected (I) or recovered (R) (\cite{kermack1927contribution}). An agent is fully described by its group membership adopting one of three values: S, I or R. This model can be written using the ordinary differential equations (1) (\cite{kermack1927contribution}). It implies a deterministic model with continuous time scale. 

\begin{gather}
    \nonumber \frac{dS}{dt} = - \beta S I \\
    \nonumber \frac{dI}{dt} = \beta S I - \gamma I\\
    \frac{dR}{dt} = \gamma I
\end{gather}

\noindent The basic reproduction number is defined by Equation 2:
\begin{equation}
    R_0 = \frac{\beta}{\gamma}
\end{equation}

\subsubsection{SIRS Model} \hfill \break
In turn, the SIRS model is another extension of the previously discussed models. Again, the disease splits the population up into 3 stages: Susceptible (S), Infected (I) and Recovered (R). However, in this model recovered agents now return to the susceptible category with a immunity loss rate (waning immunity rate) $\alpha$. As a consequence, agents can get infected multiple times as time goes on. As recovered (R) agents are immune, infection is only possible when an agent is susceptible (S) by an infected (I) agent (\cite{KupermanSIRS}). If $\alpha = 0$, the SIRS model reduces to the SIR model. The SIRS model can be expressed by the equations in Equation 3.

\begin{gather}
    \nonumber \frac{dS}{dt} = - \beta S I  + \alpha R \\
    \nonumber \frac{dI}{dt} = \beta S I- \gamma I\\
    \frac{dR}{dt} = \gamma I - \alpha R
\end{gather}

\subsection{Agent-Based Models}
As briefly mentioned before, a drawback of compartmental models (CMs) is that they are limited in the complexity that they can represent. In general, CMs and AMs often produce similar results, though AMs are able to track individuals throughout time and can thus produce extra results (\cite[p.6]{gallagher2017comparing}). Agent-based models (AMs) can encode extra behaviour by characterizing each agent with a set of variables, identifying their state (as opposed to only 1 variable). Accordingly, it revolves around the conceptualisation and analysis of stylised - and minimalistic - models that capture specific mechanisms at work. We will measure the macro behaviour of the system that emerges from the micro behaviour of agents. An agent-based model also exists for the SIR framework. \citeauthor{gallagher2017comparing} defines an agent-based model within the SIR framework as in Equation 4. $a_n (t)$ indicate dynamic agents for $n=1,2,...,N$, along with a forward operator which updates the state of the agents from one step to the next (\cite[p.8]{gallagher2017comparing}). $Bn$ denotes the Bernoulli distribution.

\begin{equation}
    a_n(t+1) = \left\{
    \begin{array}{ll}
        a_n (t) + Bn(\frac{\beta I(t)}{N}) & \mbox{if } a_n (t) = 1 \\
        a_n (t) + Bn(\gamma ) & \mbox{if } a_n (t) = 2 \\
        a_n (t) & \mbox{otherwise}
    \end{array}
\right.
\end{equation}

\subsection{Gillespie Algorithm}
The Gillespie algorithm is one of the most important stochastic simulation algorithms. It simulates Markovian processes where objects change status. For the system in a given state, the algorithm computes the time to the next event $\tau$ (probability density function given in Equation 5) and what that event will be (probability of event v given in Equation 6) (\cite{Wearing2014}). This technique is particularly efficient for processes where only the number of objects in each status matter, such as a disease spread in a well-mixed, mass-action population (\cite{kiss2017mathematics}). 
\\
\begin{equation}
    f(\tau)=(\sum_{i}a_i) exp(- \tau \sum_{i}a_i)
\end{equation}
\begin{equation}
    P(Event = v) = \frac{a_v}{\sum_{i}a_i}
\end{equation}
with $a_i$ the event rate for possible events, such as $i=$ {birth, transmission, recovery, death,...}.

\begin{algorithm}[h]
\SetAlgoLined
\KwResult{Global model trained}
\textbf{Initialize: } $\boldsymbol{w^0}$ \;
 \For{\textup{each global round $t=0,1,\dots$}}{
    \For{\textup{each client $k \in \{1, \dots K\}$}}{
        $\boldsymbol{w_k^{t+1}} \gets$ \texttt{ClientUpdate}($k,\boldsymbol{w^t}$)
    }
    $\boldsymbol{w^{t+1}} = \sum_{k=1}^K \frac{n_k}{n} \boldsymbol{w^{t+1}_k}$\;
 }
 \caption{\texttt{FedAvg} (Federated Averaging)}
 \label{fedavg}
\end{algorithm}

\section{Methods}

To evaluate how disease dynamics change depending on the underlying structure, the same epidemics model (e.g. SIR) will be ran on two or more (networkx) graphs. Furthermore, different models can be used on the same graph to evaluate the underlying structure of the network. For our analysis, we will be using both synthetic graphs and real-world graphs. During the lectures, a lot of interesting measures were discussed. To keep the project compact, the most relevant/interesting metrics for our purpose were identified.



\subsection{Networks}
\subsubsection{Synthetic Graphs}
\begin{itemize}
    \item \textbf{Erdős–Rényi (ER)}: An ER graph $G(n,p)$ depends on two parameters: $n$, number of nodes, and $p$, probability that a given edge (i, j) is present. The algorithm generates random graphs by selecting pairs of nodes with equal probability and connecting them with probability p. All nodes are equally important. The degree distribution is binomial, but can be approximated by Poisson. However, this is very unrealistic: these types of networks have a very low clustering coefficient (because $p$ is a constant, random, and independent probability) and do not account for the formation of hubs.
    \item \textbf{Watts–Strogatz (WS)}: The WS algorithm produces random graphs with small-world properties, such as short path lengths, high clustering coefficients and an unrealistic degree distribution. 
    \item \textbf{Barabási–Albert (BA)}: The Barabási-Albert (BA) algorithm generates random, scale-free networks using a preferential attachment mechanism. This creates some nodes with relatively unusually high degrees (\textit{hubs}). The degree distribution is more realistic than the ER and WS graphs, as it follows a power-law.     
\end{itemize}

\subsubsection{Real-World Graphs}
\begin{itemize}
    \item \textbf{Facebook Friendships}: Nodes represent Facebook users and edges represent their friendship relations (\cite{JulianMcAuley2013}).
    \item \textbf{Sex Escorts}: Nodes represent sex-sellers (females) and sex-buyers (males) and edges represent a sexual encounter between the two (\cite{Rocha2010InformationDS}).
    \item \textbf{Contact Tracing Graph}: Nodes represent individuals, edges represent encounters with a distance $\leq$ 5m. Mobile phone GPS data from the town Haslemere, collected over three consecutive days (\cite{klepac2018contagion}). 
\end{itemize}

\subsection{Metrics}
\begin{itemize}
    \item \textbf{Node-degree distribution}: In a social network, node degree can be interpreted as the number of contacts. The degree distribution provides insight into the developmental or growth processes that have shaped network topology (\cite[p.121]{node}).
    \item \textbf{Scale-free network}: Real networks often show a heavy-tailed skewed node-degree distribution where, apart from a few extremely linked nodes, most nodes only have a few connections. Usually, the power-law exponent is a value between 2 and 3.
    \item \textbf{Density}: The ratio of potential connections (contacts) to all possible connections. This gives us an idea of how connected the network is. 
\end{itemize}

\subsection{Procedure}


The experiments observe the effect of input parameters (network) or model parameters (infection/recovery rate) on the disease spread. The infection rate $\beta$ can be reduced by limiting social contact and wearing masks, while the recovery rate and waning immunity rate are difficult to change using regulations. They could be improved by means of medication/vaccination, but its effects are still heavily dependent on the individual. Therefore, all these rates are quite difficult to estimate and educated guesses are used in the experiments. 
\section{Results}
\subsection{Experiment 01}
First, the influence of the basic reproduction number $R_0$ is investigated on 3 synthetic networks (\tablename ~\ref{tab:synthetic_networks}) with 1000 nodes and an average degree of 10, and a well-mixed population. The recovery rate is kept at 1, while the infection rate is varied between 0 and 0.3 (30\%).

\captionof{table}{Analysed synthetic networks}
\begin{tabular}{p{1.2cm} p{1.05cm} p{2.1cm} p{1.4cm}}
\hline
Network &$<k>$ &Scale-free &Density\\
\hline
1 (BA) & 9.95 &True (2.72)& 0.9960\%\\ 
2 (ER) & 9.86 &False (8.22)& 0.9872\%\\  
3 (WS) & 10.0 &False (10.13)& 1.0010\%\\  
\label{tab:synthetic_networks}
\end{tabular}

\begin{Figure}
\centering
\includegraphics[scale=0.5]{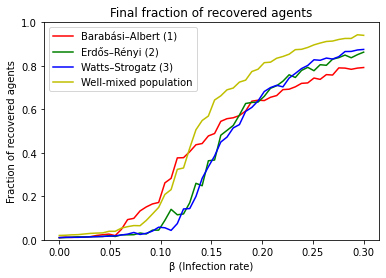}
\captionof{figure}{Epidemic scope simulation on \newline synthetic networks}
\label{fig:epidemic_scope_synthetic}
\end{Figure}

As shown in \figurename ~\ref{fig:epidemic_scope_synthetic}, a phase transition of the epidemic scope occurs when the infection rate becomes larger than $\beta = 0.10$ for networks 2, 3 and the well-mixed population. This means that the epidemic dynamics dramatically change for \textbf{non} scale-free networks changes when the basic reproduction number $R_0$ ($\frac{\beta <k>}{\gamma}$) becomes larger than 1. The same analysis was carried on three real world networks (\tablename ~\ref{tab:real_world_networks}).

\captionof{table}{Analysed real world networks}
\begin{tabular}{p{1.2cm} p{1.05cm} p{2.1cm} p{1.4cm}}
\hline
Network &$<k>$ &Scale-free &Density\\
\hline
 4 (FF) & 43.69 & True (2.51) & 1.0820\%\\ 
 5 (SE) & 4.67 & True (2.87) & 0.0279\%\\  
 6 (CT) & 6.46 & False (3.0) & 1.5490\%   
\label{tab:real_world_networks}
\end{tabular}

\begin{Figure}
\centering
\includegraphics[scale=0.5]{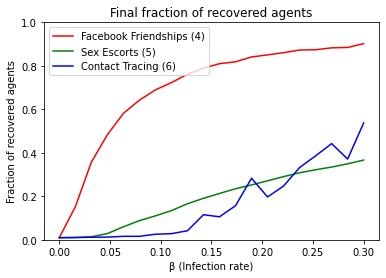}
\captionof{figure}{Epidemic scope simulation on \newline real world networks}
\label{fig:epidemic_scope_real}
\end{Figure}

As seen in \figurename ~\ref{fig:epidemic_scope_real}, the same behaviour can be observed for the contact tracing graph (6). However, for the FF (4) and SE (5) graphs, which are both scale free, no transition phase can be identified on the plot above. Even for very small values of $\beta$, a large fraction of the population gets infected and recovered.\\

This coincides with the work of \citeauthor{ShwetaBansal2007}, who pointed out that scale free networks have no epidemic threshold. As seen during the lectures, if the exponent of the power-law degree distribution is smaller than 3, the variance diverges and the epidemic threshold vanishes.

\begin{Figure}
\centering
\includegraphics[width=\linewidth]{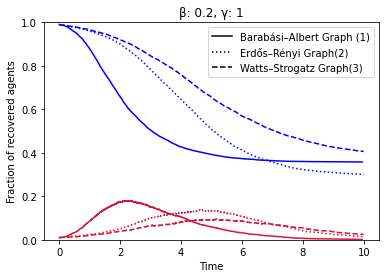}
\captionof{figure}{Network comparison}
\label{fig:network_comparison}
\end{Figure}

Scale-free networks have central hubs, nodes with a high number of connections, that can spread diseases quickly once the hub gets infected (cf. superspreaders). As seen in \figurename ~\ref{fig:network_comparison}, this allows diseases to spread faster and more intensively. For the scale-free graph (1), the curve of infections has an earlier and higher peak compared to the non scale-free networks (2) and (3), showing a smoother progression of infections.

\subsection{Experiment 02}
Secondly, the influence of two network properties on the epidemic spread is investigated: the network density and the the scale-free property of the degree distribution. In a scale-free network, \emph{superspreaders} (which correspond to the nodes with the highest degree) may have a higher impact on the epidemic spread due to their exponentially larger number of contacts present in the network (\cite{szabo2020propagation}). For this experiment, the fractions of the maximum infected agents at any time and of the final recovered agents are compared between ER and BA graphs of different sizes. The infection rate is fixed to 0.1 (10\%) and the recovery rate is fixed to 1. Further, the fraction of initial infected is set to 1\% of the total nodes. From \figureautorefname~\ref{fig:density-ir}, it can be concluded that the results for both types of graphs converge as the network density increases. Whilst there are big differences in the epidemic spreading for the two models for a density below 0.3\%, the disease spreads in an exponential manner for higher densities in both networks. For both types of graphs, a density below 0.75\% keeps the maximum relative infections below 40\% whilst showing a final fraction of recovered $\geq$75\%. The density of a network can be decreased by removing edges. In practice, this would mean (drastically) reducing each individual's social contacts. This is what measures like a home-office duty or a (more drastic) lockdown try to achieve.

\begin{Figure}
    \centering
    \includegraphics[width=\linewidth]{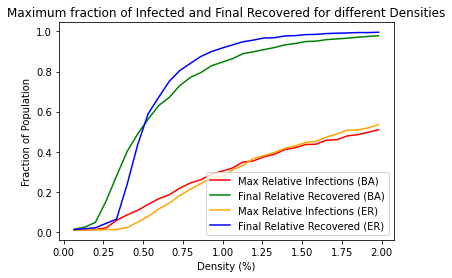}
    \captionof{figure}{Comparison of Network Densities}
    \label{fig:density-ir}
\end{Figure}

\subsection{Experiment 03}
Subsequently, this experiment tries to evaluate the impact of a measure aimed to drastically reduce social contacts like a lockdown. The introduction of such a measure is simulated on a synthetic, scale-free network (BA graph with 3.000 nodes) with an original density of 1.33\%. The measure is implemented by removing edges from the network at a given time step. The edges are removed in such a way that no node is connected to more than 5 nodes after the introduction of the measure. This is implemented by sorting the nodes' degree in a descending manner and removing all edges (which are randomly shuffled), except for the first 5. In case the node already has a degree $\leq$5, no edges are removed. In this way, the density is reduced to $<$0.33\%. After the introduction of the measure, the network is also not scale-free anymore. The infection rate and recovery rate are the same as in the previous experiment, left unchanged by the measure. Additionally, the maximum relative infections after the introduction of the measure are measured with a delay of 33\% of the remaining time steps returned by the algorithm implementation. This delay is introduced because a measure may not show an immediate effect on the maximum number of infections. The results for introducing the measure in different time steps are shown in \figurename~\ref{fig:time-ir}. At any time step, implementing such a measure can help to reduce the maximum rate of infected people. However, if the disease has already spread too much, the effects of the measure will be too little. Hence, an early introduction is recommended to reduce the maximum fraction of infected agents according to the SIR model.

\begin{Figure}
    \centering
    \includegraphics[width=\linewidth]{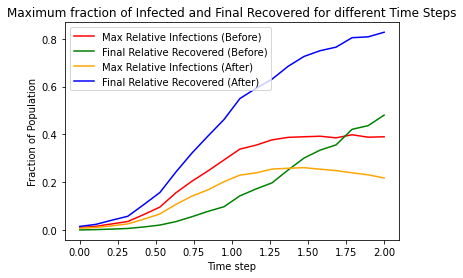}
    \captionof{figure}{Comparison of Intervention Times}
    \label{fig:time-ir}
\end{Figure}

\subsection{Experiment 04}
In all SIR simulations the number of infected agents converges to zero, independent of the input parameters. The virus slowly dies out as an agent cannot get infected twice and at a certain point in time there will not be enough new susceptible hosts left. This simplifying assumption is valid if only the initial period of an epidemic is considered. However, for a longer-term analysis, it should be taken into account that once immunity has been gained, it only lasts a limited amount of time before agents become susceptible again. Depending on the specifics of the disease more fine-grained models can be used. Therefore, the so-called SIRS model was used in order to understand the effect reinfection can have on the spread of an epidemic.\\
\begin{Figure}
\centering
\includegraphics[scale = 0.5]{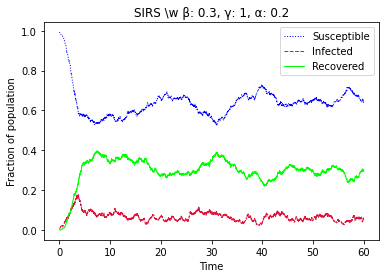}
\captionof{figure}{SIRS simulation on Contact Tracing Network}
\label{fig:sirs_real_world}
\end{Figure}

The quantitative effects of this model change (SIR $\to$ SIRS) can be seen in \figurename ~\ref{fig:sirs_real_world}. The epidemic was simulated using the same input parameters ($\beta: 0.3, \gamma: 1, \alpha: 0.2$) on different networks. Two main observations can be made: first of all, the dynamics have changed and the epidemic happens in waves now (like seen in the real world), and the ratio of infected agents spikes periodically.

\begin{Figure}
\centering
\includegraphics[width=\linewidth]{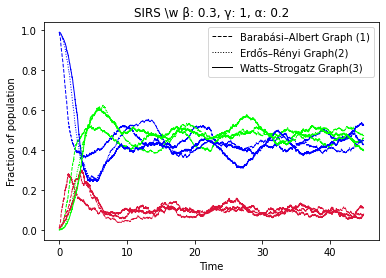}
\captionof{figure}{SIRS Simulation on synthetic Networks}
\label{fig:sirs_synthetic_networks}
\end{Figure}

Depending on the ratio of infection rate to waning immunity rate, the number of cases might show a downwards facing trend and might converge to zero. Another finding is visible in \figurename ~\ref{fig:sirs_synthetic_networks}: the underlying structure of the network becomes less relevant when studying long-term trends. We conclude that the SIR model is much more suitable for observing short-term dynamics, while more advanced models like SIRS should be used to gain long-term insights.
\section{Discussion}
The experiments demonstrate that the spread of a pandemic heavily depends on the network structure as well as on the parameters that define the disease like the infection rate. The experiments confirm that no epidemic threshold exists for scale-free graphs and very low infection rates are sufficient to lead to the spread of an epidemic. Experimenting with BA and ER graphs with different densities, it also becomes clear that the density is of great influence for the spread of the disease and the maximum infected fraction of people. For low densities, a scale-free network, such as a BA network, shows a significantly higher fraction of infected as well as final recovered people.    
Hence, mitigation strategies of the pandemic should take all those relevant parameters into consideration. For example, a lockdown reduces the number of infected people, as contacts are avoided. This is also what could be observed in real life over the past two years of the COVID-19 pandemic: Measures to reduce the infection rate ($\beta$ in the SIR model using the Gillespie algorithm) like the obligation to wear masks were combined with measures to reduce the number of contacts (edges) like a lockdown or home-office duty.

\section{Conclusion}

We were able to reproduce the finding that scale-free networks do not have an epidemic threshold using the SIR model. In addition, we could further show that the spread of the pandemic modelled by the SIR model does vary with different network densities as well as with the scale-free property. We also find that measures should, besides trying to lower the infection rate, take into account the reduction of contacts as an exemplary experiment showed that a measure that reduces both the network density and transforms the degree distribution to a scale-rich distribution can reduce the maximum concurrent infections.

\section{Contributions}
We generally tried to distribute the total workload in a manner that results in equal individual workload. In particular, Astrid focused on background research, contributing the theory and methodology. Jan and Adrian implemented the experiments on a rolling basis, incorporating the findings of the previous experiments using the EoN library, which was proposed by Jan. The usage of the Haslemere contact tracing network in addition to the real-world graphs from the lecture was proposed by Adrian. All authors revised and accepted the final version of this paper.

\printbibliography
\end{multicols}

\clearpage
\end{document}